\documentclass[preprint,aps]{revtex4}

\def\Journal#1#2#3#4{{#1}{\bf #2},#3 (#4)}
\def\prd{{ Phys. Rev.} D}
\def\pra{{ Phys. Rev.} A}
\def\prb{{ Phys. Rev.} B}
\def\npb{{ Nucl. Phys.} B}
\def\prl{ Phys. Rev. Lett.}
\def\cmp{ Comm. Math. Phys.}

\def\plb{{ Phys. Lett.} B}
\def\nuov{{ Nuovo Cim.} A}
\def\ann{{ Ann. Phys. (N.Y.)}}
\def\mpla{{ Mod. Phys. Lett.} A}

\begin{document}

\title{Duality and interacting families in models with the inverse-squared
 interaction}
\author{Ivan Andri\' c and Danijel Jurman
\footnote{e-mail address: 
iandric@irb.hr \\ \hspace*{2.2cm}
djurman@irb.hr }}
\affiliation{Theoretical Physics Division,\\
Rudjer Bo\v skovi\'c Institute, P.O. Box 180,\\
HR-10002 Zagreb, CROATIA}


\begin{abstract}

 Weak-strong coupling duality relations are shown to be present in the 
quantum-mechanical many-body system with the interacting potential 
proportional to the pair-wise inverse-squared distance in addition to the
 harmonic potential. Using duality relations we have solved the problem of
 families interacting by the inverse-squared interaction. Owing to duality,
  the coupling constants of the families are mutually inverse. The spectrum
 and eigenfunctions are determined mainly algebraically owing to $O(2,1)$
 dynamical symmetry. The constructed Hamiltonian for families and appropriate
 solutions are of hierarchical nature. 
\end{abstract}

\maketitle

\section{Introduction}

Duality is an important generalization of symmetry for studying relations
 between seemingly different theories. This symmetry is as old as the Maxwell 
equation, where it appeared for the first time. In field theory and in 
theories in higher dimensions there is a web of various dualities between 
several theories. With more degrees of freedom, duality is enlarged. With 
the exception of spin systems, there exist the Calogero \cite{C}, the
 Sutherland \cite{S} and the Moser \cite{M}
 type of rare quantum-mechanical models with duality properties.
 These are weak-strong coupling dualities, which relate various physical
 quantities depending on the constants of the interaction $\lambda$ and
 $1/\lambda$. These symmetries were found for the Sutherland \cite{G,MP1}
 model and the Calogero model without harmonic interaction \cite{AJ}. Our
 purpose here is to demonstrate that the duality of the same type operates in
 the Calogero model with harmonic interaction. Then there is an efficient
 use of duality relations to solve the old problem of interacting families of
 particles, including the inverse-squared interaction acting between
 particles belonging to different families, as well as betwen particles 
 belonging to the same family with strength that may be different for
 different families \cite{C}.
  
\indent
The system under consideration is described by the Hamiltonian
 \begin{equation}  
	- \frac{\hbar^{2}}{2m} 
	\sum_{i=1}^{N} \frac {\partial ^{2}}{\partial x_{i}^2} +
	\frac{\hbar^{2}}{2m}
	\sum_{i\neq j}^{N} 
	\frac {\lambda(\lambda-1)}{(x_{i}-x_{j})^{2}}+
	\frac{\omega^{2}}{2} \sum_{i,j=1}^{N} (x_{i}-x_{j})^{2},	
\end{equation}

which has been solved, both classically \cite{OP} and quantum-mechanically, and
has been intensively studied. This system is also related to one-matrix models
 \cite{MBIPZ,AJL} and to  two-dimensional Yang-Mills theory \cite{MP2}.
 In the large-N limit, the system possesses soliton states \cite{JPABJ} which
 are related to edge states in the quantum Hall system \cite{KAI} and the
 Chern-Simons theory \cite{ABJ}. The models are also relevant to  
 two-dimensional gravity \cite{GT} and to the Seiberg-Witten theory \cite{HP}.
 There is a remarkable connection with the physics of the black hole. The
 behavior near the horizon of the black hole is described by (1). Further
 analyzes based on (1) have been used to explore horizon states \cite{GSV,BGS}
 and shed light on black hole thermodynamics \cite{BGS}. In solving (1) we have
 restricted our attention to the case where the coupling $\lambda (\lambda-1)$
 is not strongly negative, in order to avoid the 'fall to the center'. The
 case of the strong coupling region has been analyzed using renormalization
 group techniques \cite{GR} and a new bound state appears. The Calogero
 solution was found assuming the vanishing of the wave function when
 coordinates of any two particles coincide. Such a boundary condition is
 represented by the Jastrow factor $\prod_{i<j} (x_{i}-x_{j})^{\lambda}$.
 A more general boundary condition leads to  new bound
 states \cite{GR,BMGG}.

\section{$so(2,1)$ algebra}

We shall determine the eigenvalues and eigenstates of (1) by constructing the
 representation of a spectrum generating algebra, similarly as it was done 
in Refs.\cite{AJ,GMP}. Owing to the translational invariance of the model we
 should introduce completely invariant variables \cite{P}:

\begin{equation}
	\xi_{i} \equiv x_{i}-X,\; \; \;
	X=\frac{1}{N}\sum_{i=1}^{N}x_i,\; \; \;
	\frac{\partial}{\partial \xi_{i}}\xi_{j}=\delta_{ij}-\frac{1}{N}.  
\end{equation}

The wave function of the problem will contain the Jastrow factor . 
Therefore, it is convenient to perform a similarity transformation of the 
Hamiltonian into $(\hbar =1,\;m=1)$

\[	\prod_{i<j}^{N}(x_{i}-x_{j})^{-\lambda}
	\left[
	- \frac{1}{2} \sum_{i=1}^{N} \frac {\partial ^{2}}{\partial x_{i}^2} +
	\frac{1}{2} \sum_{i\neq j}^{N} 
	\frac {\lambda(\lambda-1)}{(x_{i}-x_{j})^{2}}
	\right]
	\prod_{i<j}^{N}(x_{i}-x_{j})^{\lambda}=\]
\begin{equation}
	= -\frac{1}{2} \sum_{i=1}^{N} \frac {\partial ^{2}}{\partial x_{i}^2} +
	\frac{\lambda}{2} \sum_{i\neq j}^{N} \frac {1}{(x_{i}-x_{j})}
	\left(
	\frac{\partial}{\partial x_{i}}-\frac {\partial}{\partial x_{j}}
	\right) .
\end{equation}

Eliminating the center-of-mass degrees of freedom we obtain the generator of
 time translation

\begin{equation}
	T_{+}=
	 \frac{1}{2} \sum_{i=1}^{N} \frac {\partial ^{2}}{\partial \xi_{i}^2} +
	\frac{\lambda}{2} \sum_{i\neq j}^{N} \frac {1}{(\xi_{i}-\xi_{j})}
	\left(
	\frac{\partial}{\partial \xi_{i}}-\frac {\partial}{\partial \xi_{j}}
	\right)
\end{equation}

and the generators of scale and special conformal transformations,
 respectively, are

\begin{equation}
	T_{0}=-\frac{1}{2}
	\left(
	\sum_{i=1}^{N} \xi_{i} \frac {\partial}
	{\partial \xi_{i}}+E_{0}-\frac{1}{2}
	\right),\;\;
	 T_{-}=\frac{1}{2}\sum_{i=1}^{N} \xi_{i}^{2}.
\end{equation}

Using Eqs.(4,5) we can verify that

\begin{equation}
	[T_{+},T_{-}]=-2T_{0},\;\; [T_{0},T_{\pm}]=\pm T_{\pm}.
\end{equation}

This is the $so(2,1) \sim su(1,1)$ algebra. In the definition of the operator
 $T_{0}$ the constant $E_{0}$ is $E_{0}=\frac{\lambda}{2}N(N-1)+\frac{N}{2}$
 and $-\frac{1}{2}$ appears after removing the center-of-mass degrees of
 freedom. The important solution found by Calogero are zero-energy solutions
 $P_{m}$:

\begin{equation}
	T_{+}P_{m}(\xi_{1},\cdots,\xi_{N})=0,\;\;\;
	T_{0}P_{m}=\mu_{m}P_{m},
\end{equation}

where $\mu_{m}=-\frac{1}{2}\left(m+E_{0}-\frac{1}{2}\right)$. Calogero has
 proved that the zero-energy solutions $P_{m}(\xi_{1},\cdots,\xi_{N})$ are
 scale and translationally invariant homogeneous polynomials of degree $m$,
 written in the center-of-mass variables. Now we shall express the
 Hamiltonian (1) in terms of the generators (4) and (5). Performing the
 similarity transformation (3) on the Hamiltonian (1) and eliminating CM
 degrees of freedom we obtain

\begin{equation}
	\prod_{i<j}^{N}(\xi_{i}-\xi_{j})^{-\lambda}
	 \frac{1}{\omega}
	\left[
	- \frac{1}{2}
	\sum_{i=1}^{N}
		\frac {\partial ^{2}}{\partial \xi_{i}^2} +
		\frac{1}{2}
	\sum_{i\neq j}^{N}
	 	\frac {\lambda(\lambda-1)}{(\xi_{i}-\xi_{j})^{2}} +
		\frac{\omega^{2}}{2}
	\sum_{i=1}^{N}
		\xi_{i}^2
	\right]
	\prod_{i<j}^{N}(\xi_{i}-\xi_{j})^{\lambda} =
	-\frac{1}{\omega}T_{+} + \omega T_{-} \equiv 2L_{0}. 
\end{equation}

 The diagonalization of the Hamiltonian (1) can be achieved by
 diagonalizing $L_{0}$. In addition to $L_{0}$ we introduce raising and
 lowering operators \cite{AFF}:

\begin{equation}
	L_{\pm}=\frac{1}{2}
	\left(
	\frac{1}{\omega}T_{+}+\omega T_{-}
	\right)
	\pm T_{0},
\end{equation}

which satisfy commutation relations of the $so(2,1)$ algebra:

\begin{equation}
	[L_{0},L_{\pm}]=\pm L_{\pm},\;\;\; [L_{+},L_{-}]=-2L_{0}.
\end{equation}

 The $L$ operators are 'rotated' $T$ operators:

\begin{equation}
	L_{0}=-ST_{0}S^{-1},\;\;\; L_{\pm}=(2\omega)^{\pm 1} ST_{\mp}S^{-1}, 
\end{equation}

where  

\begin{equation}
	S=e^{-\omega T_{-}} e^{-\frac{1}{2 \omega} T_{+}}.
\end{equation}

From these equations we derive that a new set of vacuua are 'rotated'
 $T$-vacuua:

\begin{equation}
	|0,\mu_{m}\rangle = SP_{m} = e^{-\omega T_{-}} P_{m},
\end{equation}

such that

\begin{equation}
	L_{-}|0,\mu_{m}\rangle =0
\end{equation}

and

\begin{equation}
	L_{0}|0,\mu_{m}\rangle =-\mu_{-}|0,\mu_{m}\rangle .
\end{equation}

The value of the Casimir operator

\begin{equation}
	J^{2}=-L_{+}L_{-}+L_{0}(L_{0}-1)
\end{equation}

on those vacuua is

\begin{equation}
	J^{2}|0,\mu_{m}\rangle = \mu_{m}(\mu_{m}+1)|0,\mu_{m}\rangle .
\end{equation}

We shall diagonalize $L_{0}$ in terms of $T_{-}$ variables, assuming that
 eigenstates are functions $l(T_{-})$ acting on the vacuum.
From the eigenvalue equation

\begin{equation}
	L_{0}l(T_{-})|0,\mu_{m}\rangle = El(T_{-})|0,\mu_{m}\rangle ,
\end{equation}

we obtain the operator equation

\begin{equation}
	T_{-}l^{\prime \prime}+(-2\mu_{m}-2\omega T_{-})l^{\prime}+
	(2\mu_{m}\omega + \omega E)l=0,
\end{equation}

by use of Eq.(9) and the formula from Ref.\cite{AJ}, valid for the function of
 $T_{-}$:

\begin{equation}
	[T_{+},f(T_{-})]=T_{-}f^{\prime \prime} (T_{-})-2f^{\prime}(T_{-})
	T_{0}.
\end{equation}

Solutions of Eq.(18) are the well-known Laguerre polynomials:

\begin{equation}
	l\sim L_{n}^{-2\mu_{m}-1}(2\omega T_{-})
\end{equation}

with the eigenvalues $n-\mu_{m}$:

\begin{equation}
	L_{0} L_{n}^{-2\mu_{m}-1}(2\omega T_{-})|0,\mu_{m}\rangle =
	(n-\mu_{m}) L_{n}^{-2\mu_{m}-1}(2\omega T_{-})|0,\mu_{m}\rangle .
\end{equation}

In terms of the raising operators $L_{+}$, the diagonalization of $L_{0}$ is
 achieved by

\begin{equation}
	L_{0}L_{+}^{n}|0,\mu_{m}\rangle =
	 (n-\mu_{m}) L_{+}^{n} |0,\mu_{m}\rangle .
\end{equation}

This result is identical to Eq.(18) because acting on a vacuum, the raising
 operators develop a Laguerre polynomial in $2 \omega T_{-}$ owing to

\begin{equation}
	L_{+}^{n}|0,\mu_{m}\rangle=S(2\omega T_{-})^{n} P_{m}=
	e^{-\omega T_{-}} e^{-\frac{1}{2\omega} T_{+}} 
	(2\omega T_{-})^{n} P_{m}=
	L_{n}^{-2\mu_{m}-1} (2\omega T_{-}) |0,\mu_{m}\rangle .
\end{equation}    

\section{Duality}

weak-strong coupling duality relations for the Sutherland model were first
 established for the Hamiltonians in Refs. \cite{G,MP1} and used to relate the
 dynamical density correlation function for the coupling constants $\lambda$
 and $\frac{1}{\lambda}$. In Ref. \cite{AJ} duality relations were used to
 solve the problem of interacting families. From previous investigations
 in Ref.\cite{JPABJ} we
 know that duality maps particles into holes, so the wave function should
 contain the prefactor of the form

\begin{equation}
	\prod(x-z)^{\kappa}=
	\prod_{i,\alpha=1}^{N,M}(x_{i}-z_{\alpha})^{\kappa},  
	\begin{array}{l}
	i=1,...,N\\
	\alpha=1,...,M
	\end{array} ,
\end{equation}

where $z_{\alpha}$ denotes $M$ zeros of the wave function describing the
 positions of $M$ holes. Let us recall the relevant duality relations found
 in Ref.\cite{AJ}:

\[
	 T_{0}(x,\lambda)\prod(x-z)^{\kappa}=
	\left\{
	-T_{0}(z,\frac{\kappa^{2}}{\lambda})-\frac{1}{2}
	\left[
	\kappa MN+\epsilon_{0}(N,\lambda)+
	\epsilon_{0}(M,\frac{\kappa^{2}}{\lambda})
	\right]
	\right\}
	\prod(x-z)^{\kappa} ,
\]
\begin{equation}	
	 T_{+}(x,\lambda)\prod(x-z)^{\kappa}=
	\left[
	-\frac{\lambda}{\kappa}T_{+}(z,\frac{\kappa^{2}}{\lambda})+
	\frac{1+\frac{\lambda}{\kappa}}{2}\sum_{i,\alpha=1}^{N,M}
	\frac{\kappa(\kappa-1)}{(x_{i}-z_{\alpha})^{2}}
	\right]
	\prod(x-z)^{\kappa} ,
\end{equation}

where $T_{0,\pm}(z,\frac{\kappa^2}{\lambda})$ denotes an operator with the
 same functional depedence on $z_{\alpha}$ as that of the operator
 $T_{0,\pm}(x,\lambda)$ on $x_{i}$ and with the coupling constant $\lambda$
 replaced by $\frac{\kappa^{2}}{\lambda}$. Let us remind that solving of the 
 problems of the Calogero type requires just Eq.(7) to be satisfied. From
 duality relations we can construct generators for both families from $T(x)$'s
 and $T(z)$'s:

\[	
	{\cal T_{+}}=
	T_{+}(x,\lambda)+\frac{\lambda}{\kappa}T_{+}(z,\frac{\kappa^{2}}
	{\lambda})-\frac{(\lambda+\kappa)(\kappa-1)}{2}\sum_{i,\alpha=1}^{N,M}
	\frac{\kappa(\kappa-1)}{(x_{i}-z_{\alpha})^{2}},
\]

\[ 
	{\cal T_{\rm 0}}=
	T_{0}(x,\lambda)+T_{0}(z,\frac{\kappa^{2}}{\lambda}),
\]

\begin{equation}
	{\cal T_{-}}=T_{-}(x,\lambda)+
	\frac{\kappa}{\lambda}T_{-}(z,\frac{\kappa^{2}}{\lambda}) .
\end{equation}

These generators satisfy the $so(2,1)$ algebra in spite of the extension by the
 interaction term:

\begin{equation}
	[{\cal T}_{+},{\cal T_{-}}]=-2 {\cal T}_{0},\;\;\;
	[{\cal T}_{0},{\cal T}_{\pm}]=\pm {\cal T}_{\pm}.
\end{equation}

 The duality relations (26) in terms of the generators ${\cal T}_{0,\pm}$
 turn out to be a sufficient condition for solving models of the Calogero type
 with two families. The action of ${\cal T}$'s on $\prod (x-z)^{\kappa}$ is
 given by

\begin{equation}
	{\cal T_{+}}\prod (x-z)^{\kappa}=0 ,
\end{equation}

\begin{equation}
	{\cal T_{\rm 0}}\prod (x-z)^{\kappa}=-\frac{(N+M)(\kappa+1)-2}{4}
	\prod (x-z)^{\kappa} .
\end{equation}

The states on which duality relations are displayed are prefactors of the
 ground-state wave function. To diagonalize the problem of two families with
 harmonic interaction, we 'rotate' the generators (28) according to Eq.(11)
 to obtain ${\cal L}$ operators:

\begin{equation}
	{\cal L}_{0}=-{\cal S}{\cal T}_{0}{\cal S}^{-1},\;\;\;
	{\cal L}_{\pm}=(2\omega)^{\pm 1}{\cal S}{\cal T}_{\mp}{\cal S}^{-1} ,
\end{equation}

where

\begin{equation}
	{\cal S}=e^{-\omega {\cal T}_{-}} e^{-\frac{1}{2\omega} {\cal T}_{+}} .
\end{equation}

The ground state is given by

\begin{equation}
	{\cal L}_{-} {\cal S} \prod (x-z)=0 
\end{equation}

and the discrete states of ${\cal L}_{0}$ are given by

\begin{equation}
	{\cal L}_{0} {\cal L}_{+}^{n} |0,\Pi \rangle =
 	(n-\mu) {\cal L}_{+}^{n} |0,\Pi \rangle ,
\end{equation}

or in terms of the Laguerre polynomials

\begin{equation}
	L_{n}^{-2\mu_{\kappa}-1}(2\omega{\cal T}_{-}) .
\end{equation}

We interpret the ${\cal L}_{0}(x,z)=-{\cal S}{\cal T}_{0}(x,z){\cal S}^{-1}$
 as a Hamiltonian (up to similarity transformation) for two interacting
 families. After performing similarity transformation we obtain

\[
H=2 \prod_{\alpha < \beta}^{M} 
 (z_{\alpha}-z_{\beta})^{\frac{\kappa^2}{\lambda}}
\prod_{i < j}^{N}
 (x_{i}-x_{j})^{\lambda}
{\cal L}_{0}
\prod_{i < j}^{N}
 (x_{i}-x_{j})^{-\lambda}
\prod_{\alpha < \beta}^{M}
 (z_{\alpha}-z_{\beta})^{-\frac{\kappa^2}{\lambda}}=
\]
\[=\frac{1}{\omega}
	\left\{
	\left[
	- \frac{1}{2}
	\sum_{i=1}^{N}
		\frac {\partial ^{2}}{\partial x_{i}^2} +
		\frac{1}{2}
	\sum_{i\neq j}^{N}
	 	\frac {\lambda(\lambda-1)}{(x_{i}-x_{j})^{2}} +
		\frac{\omega^{2}}{2}
	\sum_{i=1}^{N}
		x_{i}^2
	\right]+\right.
\]
\[
	+\frac{\lambda}{\kappa}
	\left[
	- \frac{1}{2}
	\sum_{\alpha=1}^{M}
		\frac {\partial ^{2}}{\partial z_{\alpha}^{2}} +
		\frac {\kappa^{2}}{2 \lambda}
		\left(\frac{\kappa^{2}}{\lambda}-1\right)
	\sum_{\alpha \neq \beta}^{M}
	\frac{1}{(z_{\alpha}-z_{\beta})^{2}} +
		\frac{\omega^{2} \kappa^{2}}{2\lambda^2}
	\sum_{\alpha=1}^{M} z_{\alpha}^2
	\right]+
	\]
\begin{equation}
	\left.+
	\frac{1}{2}
	\left(1+\frac{\lambda}{\kappa}\right)
	\sum_{i,\alpha}^{N,M} \frac{\kappa(\kappa-1)}{(x_{i}-z_{\alpha})^2}
	\right\} .
\end{equation} 

The wave function of this two-family system is

\begin{equation}
	\Psi(x,z,n)\sim \prod_{\alpha < \beta}^{M} 
	 (z_{\alpha}-z_{\beta})^{\frac{\kappa^2}{\lambda}}
	\prod_{i < j}^{N} (x_{i}-x_{j})^{\lambda}
	\prod_{i,\alpha=1}^{N,M} (x_{i}-z_{\alpha})^{\kappa}
	L_{n}^{-2\mu_{\kappa}-1}(2\omega{\cal T}_{-}) .
\end{equation}

 The Hamiltonian (36) describes two families in interaction. The first family
 has particles with masses all equal to 1 and the coupling parameter $\lambda$.
In the second family, particles have masses $\frac{\kappa}{\lambda}$ and the
coupling parameter is $\frac{\kappa^2}{\lambda}$. Both physical parameters
 of the second family are of nonperturbative origin. Now it is straightforward
 to construct new families. Each new family will appear when the new prefactor
 in the zero-energy solution is introduced and new extended duality relations 
for $T_{+}$ and $T_{0}$ are established. A new ${\cal T}_{+}$ generator will be
enlarged with an additional singular interaction. These interactions have the 
same scaling dimensions as the kinetic term, so the commutation relations of
 the type $[{\cal T}_{0},{\cal T}_{+}]={\cal T}_{+}$ will remain the same even
if ${\cal T}_{0}$ is also enlarged. We construct ${\cal T}_{-}$ by adding the
 corresponding $T_{-}$'s in order to keep $so(2,1)$ algebra commutation
 relations unchanged. A new master Hamiltonian is obtained, after performing 
similarity transformation with an appropriate product of Jastrow factors from
 ${\cal L}_{0}$ which is '${\cal S}$-rotated' ${\cal T}_{0}$.

\section{Conclusion}

Using the $so(2,1)$ algebra and its generators we have first given an
 algebraic/group theoretical rederivation of known results on the Calogero
 model with harmonic interaction. It closely follows the exposition of
 de Alfaro et al. \cite{AFF}. This algebraic treatment was also used in
 the analyzis of the magnetic monopole and the vortex \cite{Jac}. We have then
 demonstrated that there exists the duality relations formulated in terms of
 the generators. This has enabled us to construct the master Hamiltonian for
 the problem of two interacting families and to construct a unique vacuum.
 This algebraic approach can be generalized to other variants of the
 Calogero-Sutherland-Moser type of models.

\section*{ Acknowledgment}

The authors gratefully aknowledge the support of the Alexander von
 Humboldt Fundation.
This work is done under contract No.0098003 supported by the Ministry of 
Science and Technology of the Republic of Croatia.


\begin{thebibliography}{99}

\bibitem{C}
F. Calogero, {J. Math. Phys.} {\bf 10}, 2191, 2197 (1969);
{J. Math. Phys.} {\bf 12}, 419 (1971).

\bibitem{S}
B. Sutherland, \Journal{\pra}{ 4}{ 2019} {1971}; \Journal{\pra}{ 5}{ 1372} 
{1972}; \Journal{\prl}{ 34}{ 1083} {1975}.

\bibitem{M}
J. Moser, {Adv. Math.} {\bf 16}, 197 (1975).

\bibitem{G}
M. Gaudin, Saclay preprint SPhT/92-158.

\bibitem{MP1}
J. A. Minahan and A. P. Polychronakos, \Journal{\prb}{ 50}{ 4236}{1994}.

\bibitem{AJ}
I. Andri\' c and L. Jonke, \Journal{\pra}{ 65}{ 034707}{2002}.

\bibitem{OP}
M. A. Olshanetsky and A. M. Perelomov, {Phys. Rep.} {\bf 71}, 313 (1981);
 {Phys. Rep.} {\bf 94}, 313 (1983).

\bibitem{MBIPZ}
M. L. Mehta, {\em Random Matrices} (Academic Press, New York, 1991); 
E. Br\'ezin, C. Itzykson, G. Parisi, and J. B. Zuber, \Journal{\cmp}{ 59}{ 35} 
{1978}; B. D. Simons, P. A. Lee, and B. L. Altshuler, \Journal{\prl}{ 48}{ 64} 
{1994}.

\bibitem{AJL}
I. Andri\'{c}, A. Jevicki, and H. Levine, \Journal{\npb}{215}{ 307}{1983}.

\bibitem{MP2}
J. A. Minahan and A. P. Polychronakos, \Journal{\plb}{ 326}{ 288}{1994}.

\bibitem{JPABJ}
A. Jevicki, \Journal{\npb}{ 376}{ 75}{1992};
 A. P. Polychronakos, \Journal{\prl}{ 74}{ 5153}{1995};
 I. Andri\'{c}, V. Bardek, and L. Jonke, \Journal{\plb}{ 357}{ 374}{1995}.

\bibitem{KAI}
N. Kawakami, \Journal{\prl}{ 71}{ 275}{1993}; H. Azuma and S. Iso, 
\Journal{\plb}{ 331}{ 107}{1994}.

\bibitem{ABJ}
I. Andri\'c, V. Bardek, and L. Jonke, \Journal{\prd}{ 59}{ 107702}{1999}.

\bibitem{GT}
G. W. Gibbons and P. K. Townsend, \Journal{\plb}{ 454}{ 187}{1999};
S. R. Das and A. Jevicki, \Journal{\mpla}{ 5}{ 1693}{1990}.    

\bibitem{HP}
E. D'Hoker and D. H. Pong, hep-th/9903068.

\bibitem{GSV}
T. R. Govindarajan, V. Suneeta, and S.Vaidya, \Journal{\npb}{ 583}{ 291}{2000}.

\bibitem{BGS}
D. Birmingham, K. S. Gupta, and S.Sen, \Journal{\plb}{ 505}{ 191}{2000};
K. S. Gupta and S. Sen, ibrd. 526, 121(2002).

\bibitem{GR}
K. S. Gupta and S. G. Rajeev, \Journal{\prd}{ 48}{ 5940}{1993}.
 
\bibitem{BMGG}
B.Basu-Mallick, P. K. Ghosh, and K. S. Gupta, hep-th/0207040.

\bibitem{GMP}
N. Gurappa, P. S. Mohanty, and P. K. Panigrahi,
\Journal{\pra}{ 61}{ 034703}{2000}.

\bibitem{P}
A. M. Perelomov, {\em Generalized Coherent States and Their Applications} (
Springer, Berlin, 1986).

\bibitem{AFF}
V. de Alfaro, S. Fubini, and G. Furlan, \Journal{\nuov}{ 34}{ 569}{1976}.

\bibitem{Jac}
R. Jackiw, \Journal{\ann}{ 129}{ 183}{1980}; \Journal{\ann}{ 201}{ 83}{1990}.
\end{thebibliography}
\end{document}